\documentclass[10pt,letterpaper,twocolumn]{article}
\usepackage[10pt,nocopyright]{sigmin}
\usepackage{listings}
\usepackage{color}
\usepackage{epsfig,makecell}
\usepackage{ifpdf}
\ifpdf
  \usepackage{epstopdf}
\fi                                  %
\usepackage[square,comma,numbers,sort&compress]{natbib}          %
\usepackage[hyphens]{url}                                         %
\usepackage[breaklinks,colorlinks]{hyperref}                      %
\usepackage[usenames,dvipsnames,table]{xcolor}                          %
\hypersetup{citecolor=blue,linkcolor=blue}                        %
\usepackage{amsmath,amsopn,amssymb,amsthm}                        %
\usepackage{thmtools}
\usepackage[toc,page]{appendix}
\usepackage{subcaption}                                           %
\usepackage{endnotes,xspace,fancyvrb,multirow} %
\usepackage[nopatch=footnote]{microtype}
\usepackage{booktabs}                                             %
\usepackage{array,underscore,relsize}                             %
\usepackage[T1]{fontenc}                                          %
\usepackage{times}                                                %
\usepackage{fancyhdr,lastpage}                                    %
\usepackage{enumitem}                                             %
\usepackage[labelfont=bf,font=small,skip=1pt,labelsep=period]{caption}           %
\usepackage[export]{adjustbox}                                    %
\usepackage{tablefootnote}                                        %
\usepackage{graphicx}
\definecolor{glred}{RGB}{178,24,24}
\definecolor{glgray}{RGB}{130,130,130}
\definecolor{glblue}{RGB}{40,80,160}

\usepackage{algorithm}
\usepackage[noEnd=true,indLines=true,commentColor=glgray,
            italicComments=true,rightComments=true]{algpseudocodex}

\newcommand{\api}[1]{\texttt{#1}}

\newcommand{\NOT}{\texttt{!}}
\newcommand{\AND}{\texttt{\&\&}}

\newcommand{\OK}{\textsc{Ok}}
\newcommand{\FAIL}{\textsc{Fail}}

\renewcommand{\NOT}{\texttt{!}\nobreak}
\renewcommand{\AND}{\nobreak\hskip.35em\texttt{\&\&}\hskip.35em\relax}

\newlength{\glgutter}
\newenvironment{glalgo}[1][1]
  {\begin{list}{}{%
      \setlength{\leftmargin}{\glgutter}\setlength{\rightmargin}{0pt}%
      \setlength{\itemindent}{0pt}\setlength{\labelwidth}{0pt}%
      \setlength{\labelsep}{0pt}\setlength{\topsep}{0pt}%
      \setlength{\partopsep}{0pt}\setlength{\parsep}{0pt}%
      \setlength{\itemsep}{0pt}}%
   \item\relax\begin{algorithmic}[#1]}
  {\end{algorithmic}\end{list}}

\newcommand{\glmark}{}
\algrenewcommand{\alglinenumber}[1]{%
  \llap{\makebox[\glgutter][c]{\glmark}}\footnotesize #1:%
  \global\let\glmark\empty}

\makeatletter
\newcommand{\ann}[2]{%
  \begingroup
  \ifx\relax#1\relax\let\algpx@beginComment\@empty\fi
  \Comment{\ifx\relax#1\relax\else#1\fi
    \ifx\relax#2\relax\else\;\textcolor{glblue}{\upshape\S#2}\fi}%
  \endgroup}
\makeatother
\newcommand{\note}[1]{\ann{#1}{}}
\newcommand{\lnote}[1]{\Statex\textcolor{glgray}{\itshape$\triangleright$\,#1}}

\algrenewcommand{\algorithmicforall}{\textbf{foreach}}
\newcommand{\comp}[1]{\,\textcolor{glgray}{\textsc{\footnotesize[#1]}}}

\usepackage{hhline}
\usepackage{pgf}
\usepackage{soul}
\usepackage{contour}
\usepackage{pifont}

\definecolor{tablegray}{gray}{0.85}

\usepackage[normalem]{ulem}
\usepackage[hang,flushmargin]{footmisc}
\usepackage{textcomp}
\usepackage{balance}
\usepackage{placeins}
\usepackage{float}

\usepackage[compact]{titlesec}
\titleformat*{\section}{\large\bfseries}
\titleformat*{\subsection}{\normalsize\bfseries}
\titleformat*{\subsubsection}{\normalsize\bfseries}
\titlespacing*{\section}{0pt}{*2.}{*1}
\titlespacing*{\subsection}{0pt}{*1.5}{*0.8}
\titlespacing*{\subsubsection}{0pt}{*1.3}{*0.8}

\definecolor{mygreen}{rgb}{0,0.6,0}
\definecolor{mygray}{rgb}{0.5,0.5,0.5}
\definecolor{mymauve}{rgb}{0.58,0,0.82}

\hypersetup{
  colorlinks,
  linkcolor={red!50!black},
  citecolor={blue!50!black},
  urlcolor={blue!80!black},
  pdftitle={UMLoad: Fast On-Demand Model Loading on Unified Memory Architecture},
  pdfauthor={Zimeng Wang, Jinrong Yang, Rong Chen}
}

\newcommand{\sys}{{UMLoad}\xspace}

\newcommand{\stitle}[1]{\vspace{0.8ex}\noindent{\bf #1}}

\newcommand{\squishlist}{
  \begin{list}{$\bullet$}{
    \setlength{\itemsep}{0pt}
    \setlength{\parsep}{3pt}
    \setlength{\topsep}{3pt}
    \setlength{\partopsep}{0pt}
    \setlength{\leftmargin}{1.5em}
    \setlength{\labelwidth}{1em}
    \setlength{\labelsep}{0.5em}
  }
}

\newcommand{\squishend}{
  \end{list}
}

\usepackage{tikz}

\usepackage{upgreek}

\begin{document}

\title{\Large \bf {\sys}: Fast On-Demand Model Loading on Unified Memory Architecture}

\author{%
  Zimeng Wang, Jinrong Yang, Rong Chen \\
  {\normalsize\itshape Institute of Parallel and Distributed Systems,} \\
  {\normalsize\itshape School of Computer Science, Shanghai Jiao Tong University}
}

\date{}
\maketitle

\frenchspacing

\begin{abstract}

On-demand model loading dominates response latency for interactive on-device AI,
yet unified memory architectures (UMAs)---now common on edge devices---should
make it cheap by placing weights in memory the GPU already sees. However,
frameworks still use the loading flow built for discrete GPUs, leaving significant inefficiencies.
This paper presents {\sys}, a system-level model-loading accelerator that
decouples a framework's logical allocations from their physical page backing.
{\sys} uses on-demand reconstruction to eliminate the weight copy while
preserving copy independence, and GPU-assisted page zeroing to accelerate
runtime allocation while preserving its security guarantee.
Implemented on Intel, AMD, and NVIDIA UMA platforms and evaluated with llama.cpp,
PyTorch, and vLLM across six models, {\sys} reduces average loading time by 36.6\%--71.9\%
over vanilla and 19.9\%--62.7\% over pipelined loaders, and cuts mean LLM-routing time to first token (TTFT)
by 51.3\%--59.3\% and model-backed tool completion latency by 10.2\%--31.6\%.

\end{abstract}

\section{Introduction}

On-device AI is moving from isolated tasks to applications that combine many
models, pairing general-purpose large language models (LLMs) with specialized ones for speech, vision,
retrieval-augmented generation (RAG), and sensing. AI use at the edge is therefore \textbf{on-demand}, in both
\emph{which} model runs and \emph{when}. Model choice is \textbf{dynamic}, based
on task requirements~\cite{hugginggpt} and cost--quality
trade-offs~\cite{routellm}, making it hard to know which models to keep in memory.
Invocation timing varies with bursty user interactions~\cite{falaki2010diversity}
and reactive agent execution~\cite{agent-xpu}, making advance loading difficult.
On memory-constrained devices, competing applications can evict model weights
and inference state, requiring restoration when the model is invoked
again~\cite{yu2026mzcache}. When the selected model is not resident,
loading enters the request's critical path.

\begin{figure}[!t]
    \includegraphics[width=\linewidth]{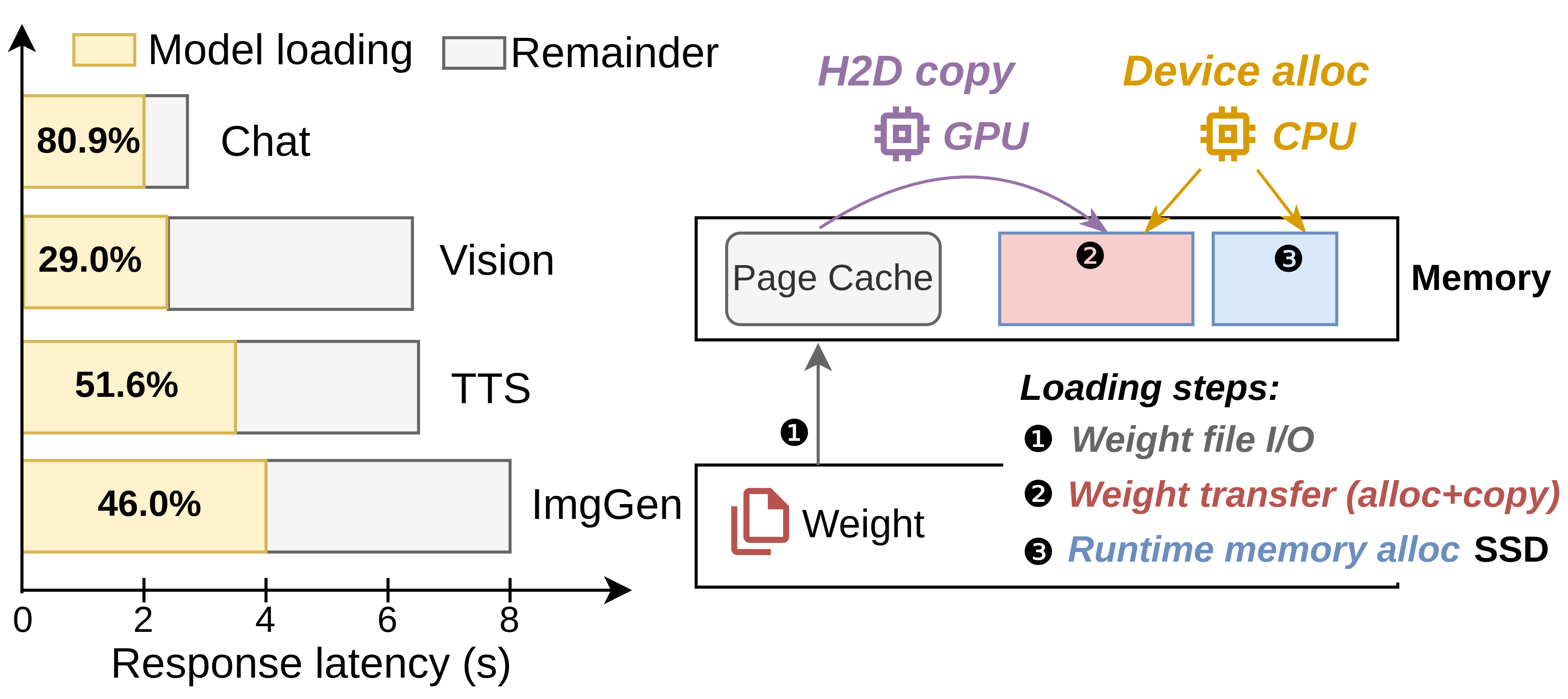}
    \caption{\small{(a) Model-loading contribution to response latency:
    end-to-end completion time for non-streaming tasks and TTFT for streaming tasks.
    Percentages indicate the fraction spent on model loading.
    Weight-file sizes are 3.7\,GB (Chat), 1.7\,GB (Vision), 4.5\,GB (TTS), and 6.5\,GB (ImgGen).
    (b) Step-by-step illustration of model loading on UMA.}}
    \label{fig:intro-loading-breakdown}
\end{figure}

Model loading can dominate cold-start {response latency} on mobile
devices~\cite{yanEdgeFlow2026}, and measurements across commercial smartphones
show substantial model-preparation delays~\cite{cotsmobile}.
We measure response latency from user input to the first user-visible result:
end-to-end completion time for non-streaming tasks and time to first token
(TTFT) for streamed text responses~\cite{agrawal2024sarathi}.
Loading costs are especially exposed when inference consists of a short subtask;
long-context prefill and lengthy generation can narrow the gap.
In the four evaluated on-demand
workloads, model loading accounts for 29.0\%--80.9\% of response latency
(Figure~\ref{fig:intro-loading-breakdown}(a), detailed in \S\ref{sec:case-study}).

\begin{figure}[!t]
    \includegraphics[width=\linewidth]{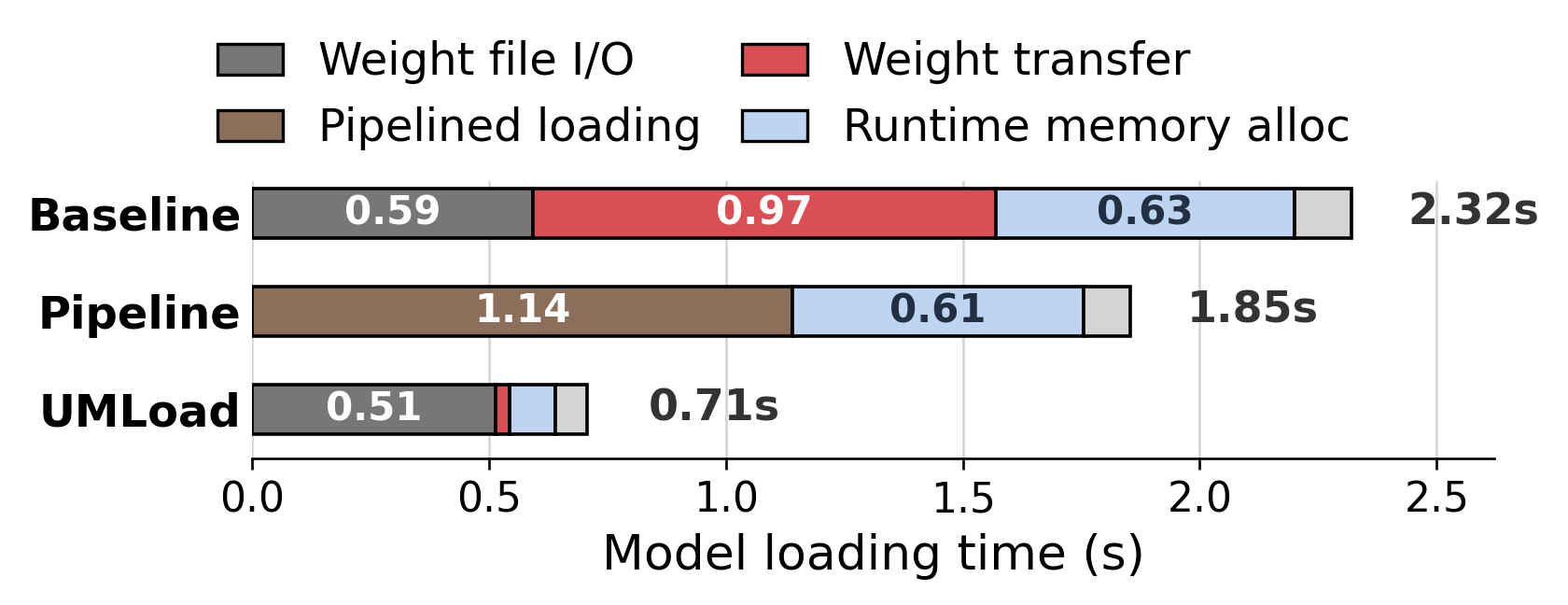}
    \caption{\small{Breakdown of Qwen3-1.7B model loading, with 3.7\,GB of weights and a 6\,GB KV-cache reservation,
	using vanilla PyTorch, PyTorch+Pipeline, and PyTorch+{\sys}. Tests use an Intel Arc Xe-LPG integrated GPU,
    with a main-memory bandwidth of 89.6\,GB/s and an SSD bandwidth of 15.75\,GB/s.}}
    \label{fig:micro-breakdown}
\end{figure}

An architectural trend offers an opening to attack this cost.
Edge devices increasingly adopt a \textbf{unified memory architecture (UMA)},
where the CPU and GPU share one physical memory pool, as in Apple M-series,
AMD APUs, Intel Core Ultra, and NVIDIA
GB10~\cite{apple-m-series,amd-ryzen-ai-max,intelcorporationIntelCoreUltra2025,nvidia-dgx-spark}.
UMA suits these workloads: it lets memory-limited devices hold larger models and
longer contexts, and in principle removes CPU--GPU transfer from loading, as the
GPU already sees weights in memory.

In practice, frameworks retain the three-step loading flow designed for
discrete GPUs (Figure~\ref{fig:intro-loading-breakdown}(b)): \emph{file I/O}
reads weights into the page cache; \emph{weight transfer} allocates GPU
buffers and performs host-to-device (H2D) copies of weights into them; and \emph{runtime allocation}
reserves memory for the key--value (KV) cache and compute buffers. On UMA, this creates
two inefficiencies (Figure~\ref{fig:micro-breakdown}). First,
\textbf{weight transfer is redundant}: it preserves buffer independence
and loader portability, but duplicates weights already in shared memory,
adding copy latency and doubling their peak memory (3.7\,GB extra for
Qwen3-1.7B). Second, \textbf{runtime allocation forces synchronous page
zeroing}: GPU allocations draw from system memory, which the OS must zero
to prevent cross-process leaks. Zeroing the 6\,GB KV cache takes 0.63\,s,
as the CPU achieves under a third of the 89.6\,GB/s memory bandwidth.

Existing solutions do not close this gap.
NVIDIA GDS~\cite{noauthor-gpudirect-nodate} streams weight files from SSD directly
into GPU memory on supported configurations. On DGX Spark (GB10), however,
GDS supports only compatibility mode, preventing use of its native
storage-to-GPU DMA path~\cite{nvidia-gds-release-notes}.
Asynchronous loading~\cite{ggmlorgLlamacpp2025,runaiModelStreamer2025,
lou-hydraserve-2026,fuServerlessLLM2024} overlaps file I/O with weight transfer,
but the copy remains exposed in our UMA measurements---file I/O is only 61\%
of the 0.97\,s transfer in Figure~\ref{fig:micro-breakdown} (``Pipeline'')---and
staging buffers add to peak memory use (\S\ref{sec:bg-existing}).
Zero-copy interfaces~\cite{cuda-memory-management} offer the right abstraction,
but deliver inconsistent performance across frameworks and platforms and require intrusive,
per-framework changes to the loader.
Besides weight transfer, synchronous page zeroing remains a runtime-allocation
bottleneck on the tested platforms (\S\ref{sec:bg-alloc-rootcause}).
UMA thus calls for a loading mechanism that
(G1) eliminates weight transfer rather than overlapping it,
thereby also avoiding the staging memory that pipelined loaders need;
(G2) accelerates synchronous page zeroing on the allocation path
while preserving its security semantics; and
(G3) generalizes across UMA platforms and integrates with
existing inference frameworks with little or no modification.

\stitle{Our approach.}
We present {\sys}, a system-level model-loading accelerator for UMA devices.
{\sys} reuses source pages on CPU to
eliminates weight transfer (G1) and uses GPU-assisted zeroing to accelerate runtime allocation (G2).
It integrates with existing inference frameworks across UMA platforms
with little or no modification (G3).

\stitle{Eliminating weight transfer transparently.}
Reusing source pages must preserve independent CPU and GPU buffers.
Direct sharing allows later CPU writes to affect GPU contents.
Unconditional ownership transfer is also insufficient: CPU contents must
remain accessible, but sources without file backing cannot be reconstructed
after handoff.
{\sys} uses on-demand reconstruction:
it transfers exclusive ownership of file-backed pages in an aligned,
huge-page layout to the GPU and reconstructs the CPU source in
separate pages only upon later CPU access.
This preserves copy independence while supporting ordinary GPU reads and writes.
The aligned, contiguous huge-page mapping preserves efficient GPU address translation.
During model loading, the CPU rarely accesses these pages afterward, so reconstruction is almost never triggered, preserving copy independence with virtually no additional memory overhead  (\S\ref{sec:weight-upload-elimination}).

\stitle{Accelerating synchronous page zeroing.}
Offloading page zeroing to the GPU shifts the bottleneck from zeroing bandwidth
to CPU orchestration: page acquisition, GPU mapping, and cache maintenance remain substantial.
\sys combines two techniques
to turn fast GPU zeroing into fast allocation.
First, it splits each allocation request into
page batches and pipelines CPU work with GPU zeroing
across batches, overlapping CPU and GPU execution.
However, CPU work can still outlast GPU execution and
leave the GPU waiting between batches. Second, it
shortens CPU stages by preparing GPU mapping resources
ahead of time to reduce per-batch mapping work and parallelizing
cache maintenance. Together, these techniques allow allocation throughput
to approach the GPU's zeroing throughput.
Throughout, pages are still zeroed before use, preserving
the security guarantee (\S\ref{sec:allocation}).

\stitle{Implementation and results.}
We implement {\sys} on Intel and AMD without framework changes, and extend
it to NVIDIA Jetson AGX Orin with minor changes to the framework (due to the closed-source CUDA runtime).
In the overall evaluation with llama.cpp, PyTorch, and vLLM across six
models, per-platform average loading-time reductions are
36.6\%--71.9\% over vanilla and 19.9\%--62.7\% over pipelined loaders.
\sys's peak memory during loading
is 1.01--1.02$\times$ the weight-file size, compared with 1.1--2.0$\times$ for pipelined and vanilla loading.
In application case studies, {\sys} improves user-perceivable response latency by reducing
mean LLM-routing TTFT by
51.3\%--59.3\% and model-backed tool completion latency by 10.2\%--31.6\%.

\stitle{Contributions.}
We summarize our contributions as follows:
\begin{itemize}[leftmargin=*,nosep]
    \item A characterization of on-demand workloads on UMA that identifies
	two inefficiencies %
	and explains why existing solutions fail to address them (\S\ref{sec:bgmot}).
	\item Two techniques for transparent model loading on UMA: on-demand reconstruction,
	which eliminates weight transfer while preserving copy independence
	(\S\ref{sec:weight-upload-elimination}), and GPU-assisted page zeroing, which
	accelerates runtime memory allocation while preserving its security semantics
	(\S\ref{sec:allocation}).
    \item An implementation across three UMA platforms (\S\ref{sec:implementation}) and
	an evaluation that demonstrates the advantage and efficacy of {\sys} (\S\ref{sec:evaluation}).
\end{itemize}

\section{Background and Motivation}
\label{sec:bgmot}
\subsection{On-demand AI workload on UMA platforms}
\label{sec:bg-uma}
On-device AI spans general-purpose LLMs and specialized models for
audio, vision, and content generation~\cite{he2019streaming,
achanta2021ondevice,fang2018nestdnn,li2023snapfusion,
jeongBandCoordinatedMultiDNN2022,yinLLMSystemService2024}.
  This diverse ecosystem naturally gives rise to on-demand workloads.
  For example, \emph{LLM routing} dispatches each query at runtime to
  the best-fit model among candidates of differing capabilities
  ~\cite{routellm}. Multi-model applications can also fuse outputs from
  multiple LLMs~\cite{mixtureofagents,llmblender,llmartini}.
  Similarly, \emph{agentic tool-calling} allows an orchestrator LLM to invoke these specialized
  models as tools and collect their outputs to compose a
  response~\cite{hugginggpt}. Generating calls to model APIs is also studied
  by Gorilla~\cite{gorilla}.
Cloud models can decompose requests into simpler subtasks executed by
on-device models~\cite{narayan2025minionscostefficientcollaborationondevice}.
On-device inference systems such as Ghidorah target single-request
execution~\cite{wei2025ghidorah}, while specialized agent subtasks motivate
the use of smaller language models~\cite{belcak2025smalllanguagemodelsfuture}.

\begin{figure}[tbp]
\centering
\captionsetup{width=\linewidth}
\includegraphics[width=\linewidth]{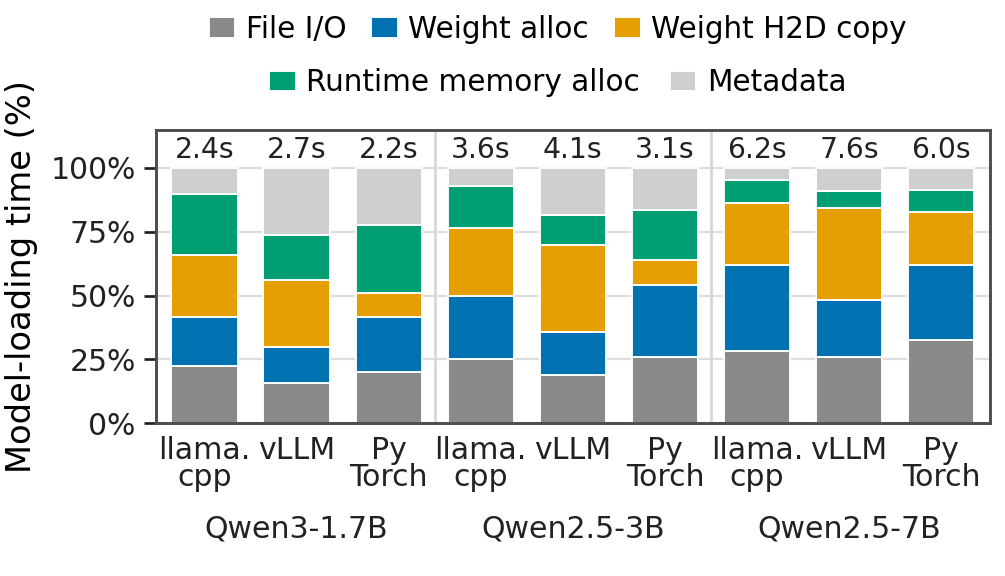}
\caption{\small{Model-loading breakdown across frameworks and models.
Runtime memory reservation is fixed at 4~GiB.}}
\label{fig:load-stage-breakdown}
\end{figure}

Edge devices have limited memory capacity and bandwidth, with other
applications further reducing the space available for resident models and
loading buffers. To use this memory efficiently, many edge platforms adopt
a \emph{unified memory architecture} (UMA), where the CPU and GPU share
one physical DRAM pool~\cite{wei2025ghidorah}. For example, Jetson Orin NX
integrates both processors with 8--16\,GB of shared LPDDR5 at
102.4\,GB/s~\cite{nvidia-jetson-orin-nx}. Besides making more memory
available to the GPU, UMA can possibly eliminate CPU--GPU copies in various workloads.

On the platforms we study, driver-managed buffer objects (BOs) record GPU
buffers' sizes and physical backing, which is mapped to GPU virtual addresses
(GVAs).
A BO can be backed by either private, driver-owned pages allocated for
GPU use or pages imported from an existing CPU buffer.
These pages are managed separately and often prepared synchronously at allocation time despite UMA's
shared main memory. Conventional AI workloads allocate BOs with synchronously prepared physical backing
and receive their associated GVAs, which the framework
uses in subsequent data transfers and GPU operations.

\subsection{Model Loading on UMA}
\label{sec:bg-loading}

\begin{algorithm}[tbp]
\caption{A common model-loading procedure.}
\label{alg:model-loading}
\small
\begin{glalgo}[1]
\Function{LoadModel}{model}\comp{framework}
  \State \lnote{Weight-file I/O}
  \State fileMap = \api{mmap}(model.file, flags)
  \State \lnote{Weight transfer}
  \State gpuMem = \api{DeviceAlloc}(\ldots)
  \ForAll{weight ranges (src, len) \textbf{in} fileMap}
    \State dst = \Call{PlanDeviceAddr}{gpuMem, len}
    \State \api{H2DCopy}(dst, src, len)
  \EndFor
  \State \lnote{Runtime memory allocation}
  \ForAll{sz \textbf{in} \Call{PlanRuntimeBuffers}{model}}
   \State \api{DeviceAlloc}(sz) %
  \EndFor
\EndFunction
\end{glalgo}
\end{algorithm}

\stitle{Loading stage breakdown.}
Algorithm~\ref{alg:model-loading} summarizes the common loading
path of the evaluated llama.cpp~\cite{ggmlorgLlamacpp2025},
PyTorch~\cite{pytorchPyTorch},
vLLM~\cite{kwonEfficientMemoryManagement2023}, and
OpenVINO~\cite{intelcorporationOpenVINODocumentation} loaders.
Despite differences in scheduling and allocation granularity,
they share three stages.

\ding{182} \emph{Weight-file I/O} accesses weights through file
mappings, represented by \texttt{mmap}%
~\cite{safetensors-mapped-loading,openvino-mapped-weights}.
Accessing mapped ranges brings their contents into the system
page cache and supplies the host addresses used for weight transfer.

\ding{183} \emph{Weight transfer} allocates GPU memory through
\texttt{DeviceAlloc} and initializes it through \texttt{H2DCopy},
using mapped weights directly or through host staging buffers.
Although allocation may be bulk or incremental, each weight
tensor typically occupies a contiguous GPU virtual-address range,
determined by \texttt{PlanDeviceAddr}.
The first use of this range after allocation is as an H2D
destination, before GPU computation consumes its contents.
Thus, the framework needs the destination address early, but
physical backing can be deferred until H2D reveals the source
pages. On UMA, eligible source pages can then directly back
the destination range, avoiding separate weight-page allocation
and copying.

\ding{184} \emph{Runtime memory allocation} then prepares memory
for inference, primarily the KV cache and compute buffers.
Their sizes depend on execution configuration and framework
planning, rather than weight-file contents alone.
\texttt{PlanRuntimeBuffers} determines these requests, which
invoke the same \texttt{DeviceAlloc} interfaces.
The driver learns the actual allocation sizes only when these
requests arrive, limiting advance preparation based solely
on preceeding stages.

Figure~\ref{fig:load-stage-breakdown} further quantifies these costs across three LLMs
and three frameworks on a representative UMA
platform. Across all nine cases, weight transfer and runtime memory allocation
dominate model loading. Device allocation alone, including weight allocation
and runtime memory allocation, consumes 28.4--48.2\% of model-loading latency.
Together, they account for
1.8--3.7$\times$ the file-I/O time.
The gap is larger for smaller models because
runtime memory allocation is relatively stable as model size grows.

\begin{figure*}[tp]
\begin{minipage}[t]{\columnwidth}
\vspace{0pt}
\centering
\captionsetup{width=\linewidth}
\includegraphics[width=\linewidth]{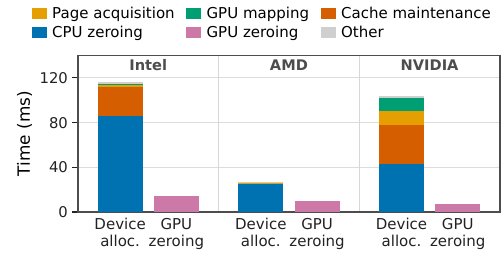}
\caption{\small{Device allocation breakdown and GPU zeroing time for
1~GiB on three UMA platforms. Allocation uses
\texttt{zeMemAllocDevice}, \texttt{hipMalloc}, and
\texttt{cudaMalloc}, respectively. GPU zeroing uses userspace drivers' native
memory-fill APIs on warmed allocations.}}
\label{fig:motivation-allocation-overheads}
\end{minipage}
\hfill
\begin{minipage}[t]{\columnwidth}
\vspace{0pt}
\centering
\includegraphics[width=\linewidth]{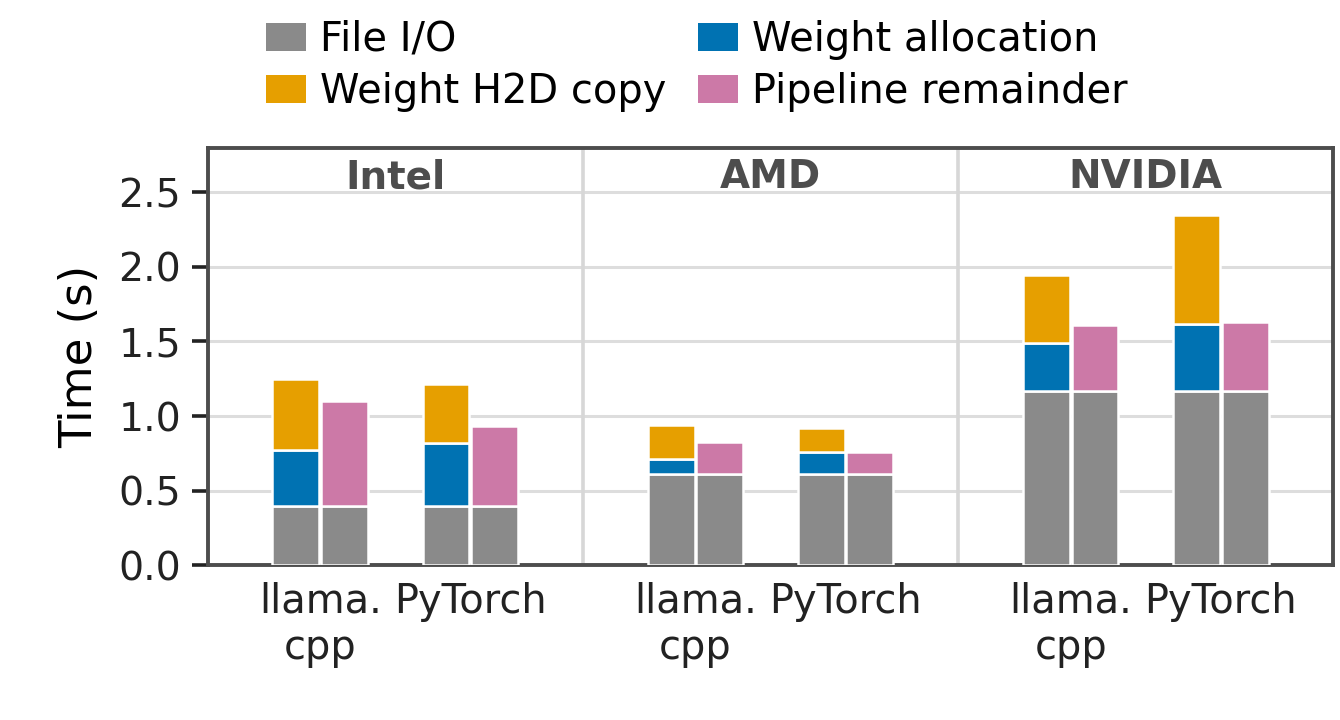}
\caption{\small{Weight-loading costs of Qwen3-1.7B on three UMA platforms.
Each pair shows vanilla loading (left) and pipelined loading (right).
The pipeline remainder is pipelined wall time minus file-I/O time,
showing work not hidden by I/O.}}
\label{fig:motivation-pipeline-overlap}
\end{minipage}
\end{figure*}

\stitle{API overhead analysis.}
\label{sec:bg-alloc-rootcause}
To explain the slow model loading, we break down the overheads of the two
GPU memory-preparation APIs involved in stages \ding{183} and \ding{184}.
First, \textbf{H2D copy} incurs work beyond GPU execution.
Pageable sources can require pinning, mapping, or
staging before the GPU can access them. Our Intel microbenchmark shows that
synchronous overheads before GPU copy execution account for
33.9\% of the API overhead for a 512~MiB H2D transfer, therefore merely using asynchronous H2D copy APIs
still blocks model loading.
Second, \textbf{device allocation} incurs substantial CPU zeroing. Figure~\ref{fig:motivation-allocation-overheads}
breaks down native allocation latency for a 1~GiB request on three UMA platforms.
CPU zeroing accounts for 41--96\% of allocation time in these cases.
GPU zeroing (estimated using a memory-fill operation),
however, is 2.5--6.4$\times$ faster than CPU zeroing.
The measured Intel and NVIDIA allocation paths also perform CPU cache
maintenance after zeroing to manually handle coherency.
The remainder includes physical-page acquisition, GPU mapping,
and other driver/runtime work.
NVIDIA's remainder is additionally expensive, because it handles
page mappings and metadata at base-page granularity on our platform.

\subsection{Current Solutions and Their Limitations}
\label{sec:bg-existing}

\stitle{Reducing model loading.}
Quantization reduces the weight representation size~\cite{lang-comprehensive-2024,
frantar2022gptq,lin2024awq,yanEdgeFlow2026}, while contextual sparsity enables
selective weight loading~\cite{llmflash}. Pruning reduces the number of nonzero
weights~\cite{frantar2023sparsegpt}; reducing file I/O additionally requires
a representation and loading path that omit those zeros.
Other approaches retain resources in pools and caches~\cite{zhuTangram2025,cuda-stream-nodate}
or optimize prefetching and eviction during loading~\cite{author-accelerating-nodate}.
These techniques reduce transferred data or exploit reuse and overlap.
They can complement \sys's elimination of redundant weight copies and
acceleration of fresh-page zeroing.

\stitle{Bypassing weight transfer.}
Drivers offer APIs that help to achieve zero-copy or even skip weight transfers between GPU and CPU on UMA platforms.
Host-registration APIs expose existing host buffers to the GPU
(e.g., \texttt{cudaHostRegister}), while shared-allocation APIs
provide memory accessible to both processors
(e.g., \texttt{cudaMallocManaged})~\cite{cuda-memory-management}.
Finally, NVIDIA GPUDirect Storage (GDS) reads files directly into
GPU buffers on supported configurations, bypassing host
staging~\cite{noauthor-gpudirect-nodate}.
However, adopting these mechanisms requires changes to
framework loading paths.
In addition, shared allocation still incurs slow zeroing on
all tested UMA platforms.
For writable private source mappings, host registration can incur synchronous copy-on-write during registration
(\S\ref{sec:eval-cr}).
Finally, on DGX Spark, GDS supports only compatibility mode, preventing use
of its native storage-to-GPU DMA path~\cite{nvidia-gds-release-notes}.

\stitle{Overlapping.}
Existing systems hide loading latency by overlapping it with
independent computation or communication~\cite{liuCSGO2025,
liPrimaCpp2026,jiaActiveFlow2025,li-tpi-llm-2025,yanEdgeFlow2026,
wang-tz-llm-2026}
Others overlap stages within loading itself, including
file I/O, weight allocation, and H2D
copies~\cite{lou-hydraserve-2026,fuServerlessLLM2024,
ggmlorgLlamacpp2025,runaiModelStreamer2025}.
However, in our measured UMA cases (Figure~\ref{fig:load-stage-breakdown}),
allocation and copying can exceed file I/O, leaving weight transfer on the critical path
and preventing the pipeline from reaching the file-I/O bound.

We examine two pipelined model loaders: llama.cpp's asynchronous loader~\cite{ggmlorgLlamacpp2025} and PyTorch's Run:ai Model Streamer~\cite{runaiModelStreamer2025},
on three UMA platforms (detailed in Table~\ref{tab:implementation-platforms}) with Qwen3-1.7B,
and their pipeline remainders are shown in Figure~\ref{fig:motivation-pipeline-overlap}.
On Intel, llama.cpp and PyTorch leave remainders of 0.70~s and 0.53~s,
respectively. The overlap is naturally limited because weight transfer already
exceeds file I/O in both frameworks (averaging 0.83~s versus 0.40~s).
Across AMD and NVIDIA, transfer takes only 0.31--0.33~s against 0.61~s of reads.
However, llama.cpp and PyTorch still leave
average remainders of 0.33~s and 0.31~s, respectively.

The two loaders expose different costs. llama.cpp allocates weight
memory and prepares upload buffers before the pipeline starts, leaving 0.13~s
on AMD and 0.41~s on NVIDIA not overlapped. It also issues
H2D copies on the same thread as file I/O, leaving API submission
work on the critical path and adding 0.085~s on AMD.
PyTorch uses multithreading to overlap
both weight allocation and H2D copies with file I/O, but staging-buffer
preparation and remaining exposed work still cost 0.15~s on AMD
and 0.47~s on NVIDIA.

Using a pipeline to reach this bound is not free, either. The staging buffer that lets Run:ai
reach high pipeline efficiency scales with the model, growing from 3.7\,GB at 1.7B to
10.4\,GB at 7B. The llama.cpp asynchronous loader requires fewer staging buffers,
but the minimum buffer size is still determined by the model's largest tensor (about 10\% of the total weight size).
Finally, neither evaluated pipeline accelerates synchronous zeroing of
fresh runtime pages.

\section{\sys Design}
\label{sec:design}

\begin{figure}[tbp]
\centering
\includegraphics[width=\linewidth]{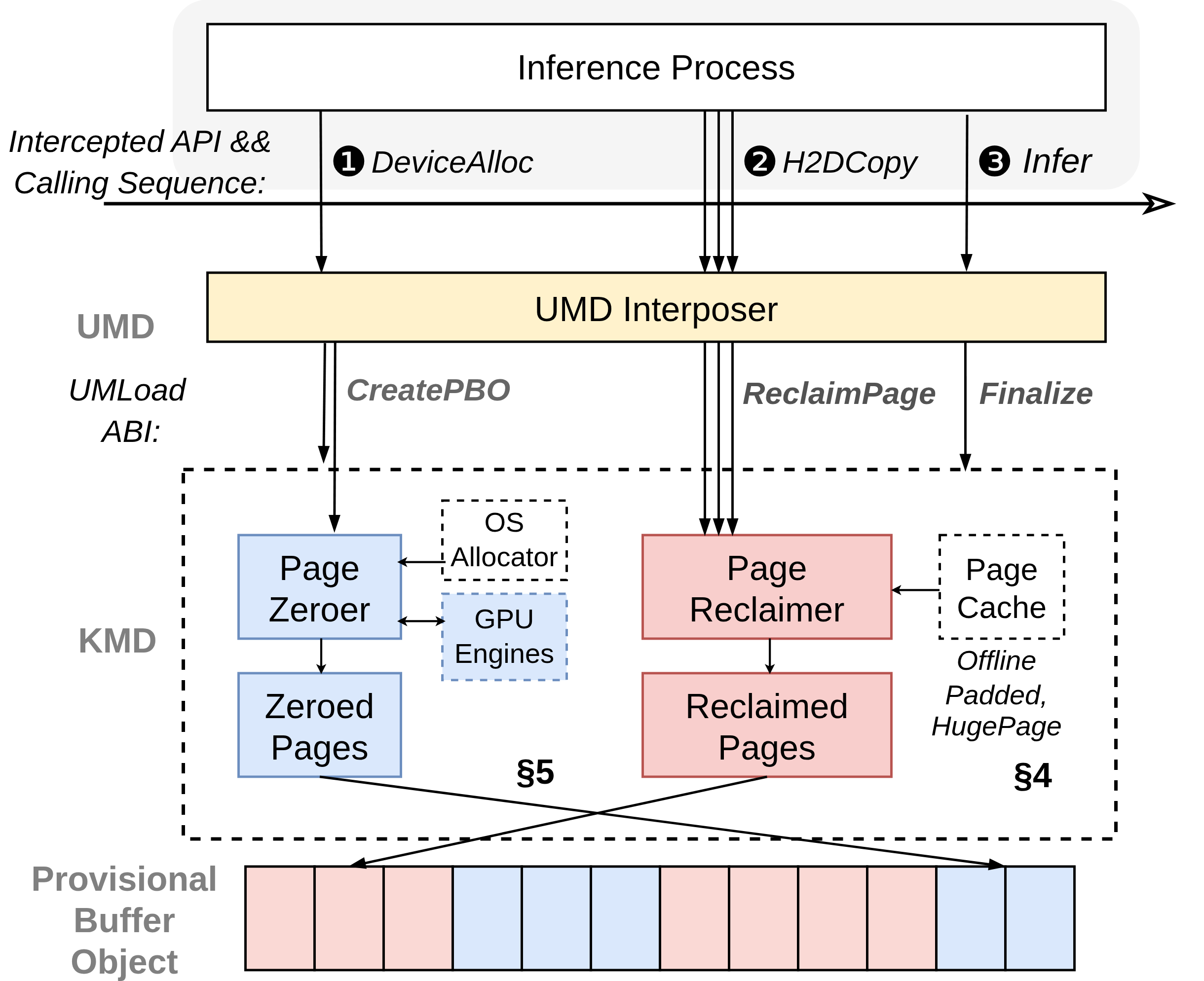}
\caption{\small{\sys architecture and interposing workflow.}}
\label{fig:workflow-lifecycle}
\end{figure}

\sys is a driver-level system that transparently accelerates model loading on UMA
platforms by eliminating redundant weight transfer (GPU weight allocation and H2D copying) and accelerating
runtime memory allocation.
\sys's transparency goal is to require no framework-side code changes and have no effect on GPU inference.
Figure~\ref{fig:workflow-lifecycle} shows
its components across the user-mode driver (UMD) and kernel-mode
driver (KMD).

\stitle{Architecture.}
Figure~\ref{fig:workflow-lifecycle} provides an overview of \sys. It spans the
user-mode driver (UMD) and the kernel-mode driver (KMD), and requires no changes
to framework code on supported paths.
The \emph{Page Reclaimer} (\S\ref{sec:weight-upload-elimination})
implements On-demand Reconstruction. Its \texttt{ReclaimPage} API
identifies eligible page-cache pages through H2D source ranges and
reclaims them with exclusive ownership as the destination buffer's
physical memory. These \emph{reclaimed pages} replace weight transfer.
Later CPU accesses reconstruct
the source contents from the backing file in separate physical pages.
The \emph{Page Zeroer} (\S\ref{sec:allocation}) accelerates runtime
memory allocation by offloading page zeroing to the GPU. It accelerates the allocation of
\emph{private pages}, zeroed system pages managed by the driver,
for buffer ranges not covered by page reclamation.

\sys uses a \emph{provisional buffer object} (PBO) to decouple a
framework-visible GPU allocation from its physical backing. A PBO exposes the
handle, size, and GPU virtual address (GVA) of an ordinary GPU buffer object
(BO), but leaves its backing open for construction from either reclaimed pages
or private pages. Internally, a PBO maintains an array of page slots and an
ownership bitmap that distinguishes the two: an empty slot is \textsc{Unbound},
and a populated one is \textsc{Reclaimed} or \textsc{Private}. Both classes
reside in UMA system memory and differ only in origin and release path. One PBO
can therefore combine several reclaimed and private ranges under a single
framework-visible GVA and lifetime.
Before forwarding a PBO for GPU execution, \sys completes its physical backing
by filling the remaining \textsc{Unbound} slots with \textsc{Private} pages
supplied by the Page Zeroer. The PBO then behaves as an ordinary GPU buffer object
in any subsequent GPU operation.

The \emph{UMD interposer} interposes driver API calls issued by the framework.
It patches the corresponding UMD logic and invokes \sys's ABIs in the KMD in the following ways.
\ding{182} As model loading starts, the framework issues device allocation API
calls for model tensors. The interposer turns each call into a \texttt{CreatePBO} call
and sends the allocation size to the Page Zeroer to begin
preparing private pages (\S\ref{sec:allocation}).
\ding{183} On an H2D copy into an open PBO, the interposer issues
\texttt{ReclaimPage} to resolve and reclaim the source pages.
Successful reclamation installs them in the destination range and
suppresses the original copy. Otherwise, \sys finalizes the PBO
as needed and follows the ordinary copy path.
\ding{184} A PBO that remains open is finalized before its first
consuming operation other than an H2D copy. \sys interposes on all
driver APIs with kernel arguments to check cases in which frameworks use
pointers directly. Supporting the remaining uncommon indirect usage requires
replayable GPU page faults.

\section{On-demand Reconstruction}
\label{sec:weight-upload-elimination}

\subsection{Eliminating Weight Allocation and H2D Copy}
Weight transfer comprises two operations: allocating GPU buffers for weights and populating them with host-to-device (H2D) copies.
On UMA, the source pages already reside in shared physical memory.
The straightforward approach
is therefore to let the GPU directly access the pages backing the H2D copy source,
eliminating redundant physical-page allocation and the H2D copy.

First, \sys replaces
intercepted allocation calls with \texttt{CreatePBO}, which reserves
GPU addresses while postponing physical-page allocation.
On the loading paths represented by Algorithm~\ref{alg:model-loading},
H2D copies initialize weight buffers before GPU computation uses them.
The PBO's physical backing can therefore remain open until H2D calls
supply the source addresses and byte ranges.
Second, when an H2D call targets an open PBO,
\sys invokes \texttt{ReclaimPage} to obtain the physical pages backing the source range, faulting
them in if necessary, and installs them as the corresponding
destination range's backing. Before the PBO's first consuming operation
other than an H2D copy, \sys finalizes it and establishes GPU page-table
entries mapping these pages at the framework-selected destination
addresses, allowing the GPU to directly access their contents.
The completed PBO then supports ordinary GPU operations.

Direct page mapping requires the source and destination to have the
same in-page offset, whereas H2D copy arguments may have arbitrary
byte offsets. To match these offsets, during a one-time preparation
run, \sys records
each weight tensor's framework-selected GPU VA.
It then rewrites the weight file using \textsc{Padding}: given
the current absolute file offset $c$ and a recorded destination
$d$, it inserts $(d-c)\bmod P$ padding bytes, where $P$ is the base-page
size. This matches their in-page offsets without changing the
framework's GPU layout.
Subsequent loads validate the offset relationship and follow the
ordinary copy path as a fallback.

\begin{algorithm}[!t]
\caption{Pseudocode for On-demand Reconstruction.}
\label{alg:page-cache-reclaim}
\small
\begin{glalgo}[1]

\Statex
\lnote{Prepare weights with \textup{\textsc{Reorder}} and \textup{\textsc{Padding}},}
\lnote{using huge-page ($G$) alignment; enable system THP.}
\Function{InterposedH2D}{dst, src, len}\comp{umd}
  \State pbo = \Call{FindOpenPBO}{dst, len}
  \If{pbo \AND \Call{ReclaimPage}{pbo, dst, src, len}}
    \State \Return \note{original H2D copy suppressed}
  \EndIf
  \If{pbo}
    \State \Call{Finalize}{pbo}
  \EndIf
  \State \Return \Call{OriginalH2DCopy}{dst, src, len} \note{fallback}
\EndFunction

\Statex
\Function{ReclaimPage}{pbo, dst, src, len}\comp{kmd}
  \lnote{1. Identify reconstructible source pages}
  \State s = \Call{EligibleSource}{pbo, dst, src, len}\label{line:reconstruct-identify-start}
  \If{\NOT s}
    \State \Return \FAIL\label{line:reconstruct-identify-end}
  \EndIf
  \If{\NOT \Call{LockAndCheck}{s}}
    \State \Return \Call{AbortReclaim}{s}
  \EndIf

  \lnote{2. Acquire exclusive GPU ownership}
  \State \Call{UnmapCPUPTEs}{s}\label{line:reconstruct-reclaim-start} \note{retain source VMA}
  \If{\NOT \Call{WriteBackWait}{s}}
    \State \Return \Call{AbortReclaim}{s}
  \EndIf
  \If{\NOT \Call{DetachPageCache}{s}} \note{refcnt drops to 1}
    \State \Return \Call{AbortReclaim}{s}
  \EndIf
  \State \Call{InstallInPBO}{pbo, dst, s.pages}\label{line:reconstruct-reclaim-end}
  \State \Call{Unlock}{s}
  \State \Return \OK
\EndFunction

\Statex
\lnote{3. Recover huge-page GPU mappings}
\Function{Finalize}{pbo}\comp{kmd}
  \State \Call{FillUnboundSlots}{pbo} \note{private pages}
  \ForAll{r \textbf{in} \Call{MappingRanges}{pbo, G}}\label{line:reconstruct-map-start}
    \If{\Call{HugeMappable}{r, G}}
      \State \Call{MapGPU}{r, G}
    \Else
      \State \Call{MapGPU}{r, P}\label{line:reconstruct-map-end}
    \EndIf
  \EndFor
  \State pbo.open = false
\EndFunction

\Statex
\lnote{4. Reconstruct on CPU access (Linux page-fault path)}
\Function{{do\_fault}}{vmf}\comp{linux, simplified}
  \State file = vmf.vma.vm\_file;\quad idx = vmf.pgoff\label{line:reconstruct-cpu-start}
  \State page = \Call{LookupPageCache}{file, idx}
  \If{\NOT page}
    \State page = \Call{ReconstructPageCache}{file, idx}
  \EndIf
  \State \Return \Call{MapCPUFilePage}{vmf, page}\label{line:reconstruct-cpu-end}
\EndFunction

\end{glalgo}
\end{algorithm}

\subsection{Preserving Copy Semantics}
\label{sec:rcl}
Directly mapping the source pages into the GPU destination leaves
the CPU and GPU accessing the same physical pages. An ordinary H2D
copy, however, allows the CPU source and GPU destination to evolve
independently. The driver cannot assume that every intercepted H2D call
transfers immutable weights. With shared pages, later CPU writes can
modify the GPU contents and vice versa.

One possible approach is to give the GPU exclusive ownership of every
source page and revoke CPU access. However, later CPU accesses must still
find valid source contents after the GPU has gained exclusive ownership of the pages.
\sys uses \emph{On-demand Reconstruction} (ODR): it gives the GPU
exclusive ownership of the source pages for ordinary GPU reads and writes,
and reconstructs the CPU source in separate physical pages upon later
CPU access.
If the GPU later writes to these pages (the framework may reuse the memory
for other purposes), no write-triggered copying is needed.
The design contains four steps.

\stitle{Identifying reconstructible sources.}
Whether the CPU contents can be reconstructed depends on the source's
backing. An H2D source may map page-cache pages or anonymous pages.
For page-cache pages, the backing file provides the
contents needed for later reconstruction. Anonymous pages, however,
may contain private contents absent from the file. Their contents cannot
be reconstructed from the file after the pages are handed to the GPU.
\sys therefore
reclaims only source ranges backed by the file page cache, excluding
anonymous pages such as stacks, heaps, and private CoW pages.

\sys checks eligibility when interposing H2D copy calls
(lines~\ref{line:reconstruct-identify-start}--\ref{line:reconstruct-identify-end}
in Algorithm~\ref{alg:page-cache-reclaim}).
If validation fails, \sys finalizes the PBO as needed and falls back to copying
from the original source. Since eligibility depends on the source's
backing, eligible transfers of non-weight file contents can also use
reclamation.

\stitle{Reclaiming pages with exclusive ownership.}
To give the GPU exclusive ownership, \sys follows Linux's page-cache
eviction mechanism. It revokes CPU page-table mappings, completes necessary
writeback, and detaches the source pages from the page cache before
installing them as the GPU destination's backing
(lines~\ref{line:reconstruct-reclaim-start}--\ref{line:reconstruct-reclaim-end}
in Algorithm~\ref{alg:page-cache-reclaim}).

Reclamation changes page-cache residency and can disrupt other
processes using the same file.
To limit this disruption, \sys rejects page-cache ranges already mapped into
multiple processes. If this check or ownership acquisition fails,
\sys follows the ordinary H2D copy path, finalizing the PBO as needed
while preserving previously initialized ranges.

\stitle{Recovering huge-page GPU mappings.}
\label{sec:efficient-reference-mapping}
Preserving inference performance also requires efficient address translation.
A tensor assembled from a source buffer can require more GPU page-table
entries than a normal GPU allocation, increasing translation overhead.
\sys recovers huge-page GPU mappings by combining weight-file layout
adjustments with contiguous page-cache backing.
Specifically, \sys
extends the earlier \textsc{Padding} from modulo $P$ to modulo
the large-mapping granularity $G$ for tensors larger than $G$
(2~MiB on our evaluated Intel and AMD platforms).
Corresponding file and GPU ranges now reach large-page boundaries
together, enabling large mappings within individual tensors.

Huge-page mappings also require contiguous physical backing.
\sys uses the OS's transparent huge-page (THP) support for the page cache
when available and installs huge GPU page-table entries at finalization
(lines~\ref{line:reconstruct-map-start}--\ref{line:reconstruct-map-end}
in Algorithm~\ref{alg:page-cache-reclaim}).
To further recover physical contiguity across tensors with consecutive
GPU address ranges, \sys applies \textsc{Reorder}, placing these tensors
consecutively in the weight file. Their shared boundary pages are checked
and reclaimed to preserve physical contiguity across tensors.
Padding is thus needed only between such groups.
The resulting layout can support a huge-page mapping for every complete,
$G$-aligned block within the group.

\stitle{Reconstructing the CPU source on demand.}
The CPU source's virtual memory area (VMA) remains valid after reclamation.
Since the reclaimed pages were backed by the file page cache,
Linux's normal file-fault path restores the source contents
from the backing file into separate physical pages upon subsequent CPU access
(lines~\ref{line:reconstruct-cpu-start}--\ref{line:reconstruct-cpu-end}
in Algorithm~\ref{alg:page-cache-reclaim}).
The GPU retains the reclaimed pages throughout. Thus, subsequent
writes to either buffer cannot affect the other, preserving ordinary
copy behavior without eagerly allocating and copying a second set
of pages or requiring GPU accesses to trigger faults and copies.
Section~\ref{sec:eval-cr} evaluates ODR's transfer latency,
inference performance, and overheads.

\section{GPU-Assisted Page Zeroing}
\label{sec:allocation}

\sys's Page Zeroer offloads page zeroing to the GPU to accelerate runtime
memory allocation on UMA. It pipelines CPU submission, GPU execution, and
CPU cleanup, then reduces CPU work to keep the pipeline supplied with
zeroing tasks.

\stitle{GPU zeroing.}
CPU zeroing accounts for much of runtime memory allocation
(\S\ref{sec:bg-loading}). A straightforward approach is to offload it to
the GPU. The KMD creates an execution context and keeps the zeroing
program available across requests.
It processes each request through
three sequential stages before starting the next:
\begin{enumerate}[leftmargin=*,nosep]
\item \emph{Task submission.} The CPU acquires raw physical pages,
performs required cache
maintenance to make them accessible to the device, installs GPU
VA mappings, and prepares task descriptors, command buffers,
and completion fences before submitting the task.
\item \emph{Task execution.} The GPU writes zeros to the mapped pages.
Following open-source driver implementations, the KMD builds command
buffers that issue hardware fill commands or launch a precompiled zeroing
kernel (detailed in \S\ref{sec:implementation}).
\item \emph{Task cleanup.} After GPU completion, the CPU performs required
visibility operations, releases GPU mappings and task resources,
and returns pages to the driver's allocation path.
\end{enumerate}

\stitle{CPU work after GPU offloading.}
The CPU performs task submission and cleanup.
Offloading zero stores leaves page and resource management on the CPU.
Page acquisition, GPU mapping, and cache maintenance
all grow with the request size.
On our measured NVIDIA platform, CPU-side work for a sequential
4~GiB request takes about 34.1~ms, accounting for 57\% of end-to-end latency.
About 49\% of the submission time is spent on page acquisition and 41\% on GPU mapping.
On Intel, CPU-side latency accounts for 56\% of the 179.41~ms request time,
and 83\% of the submission time is spent on cache maintenance, as the Intel iGPU's fast
non-coherent zeroing task still requires explicit CPU cache flushing.
GPU offloading therefore leaves substantial CPU-side memory management on the allocation path.

\begin{figure}[tbp]
\centering
\includegraphics[width=\linewidth]{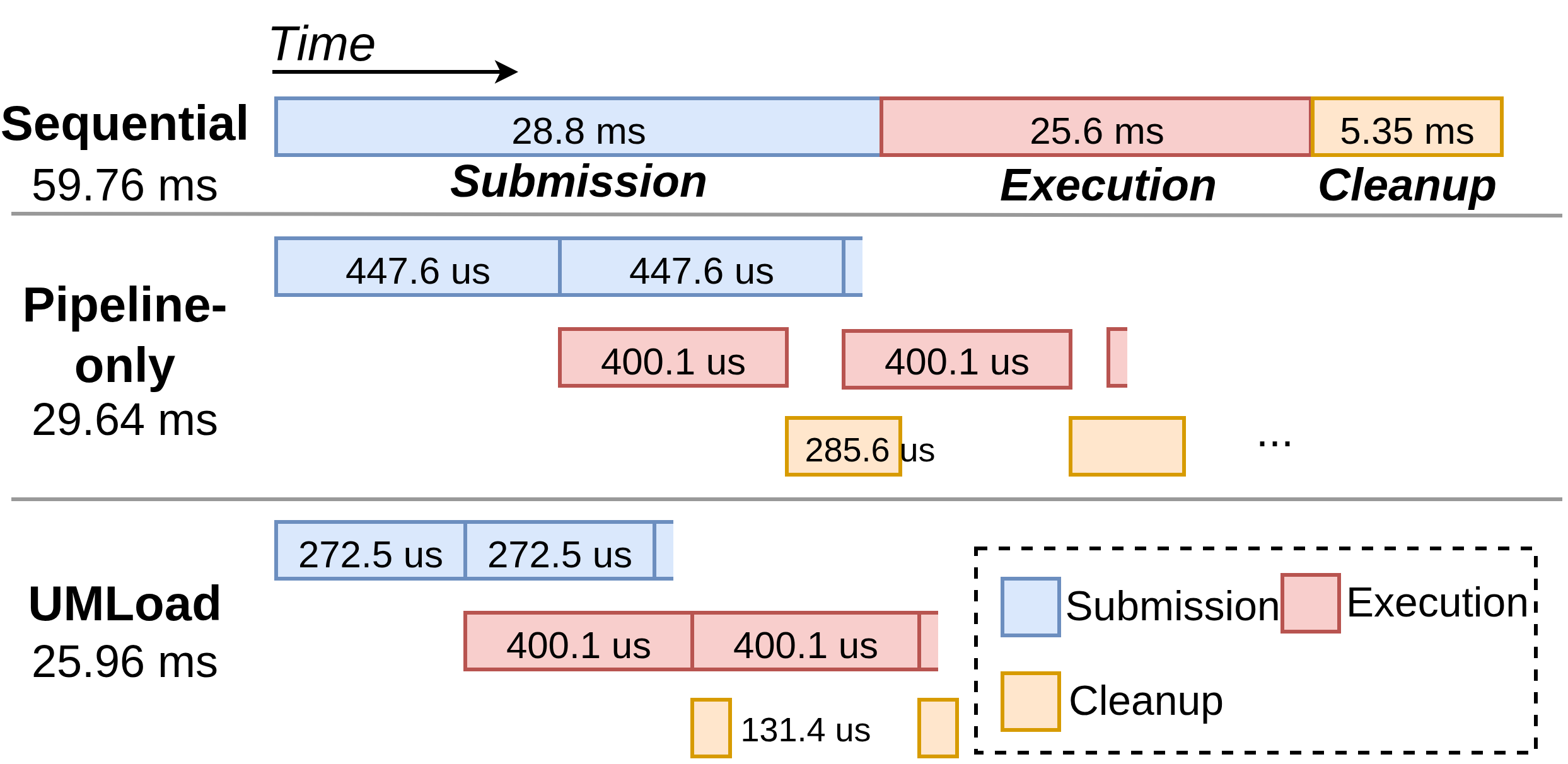}
\caption{\small{GPU-assisted zeroing of a 4\,GiB request on NVIDIA
Jetson AGX Orin.
The top row shows a sequential method, where block labels show
stage time for the whole request.
The middle and bottom rows show a pipeline-only method and \sys, respectively,
where block labels show individual 64\,MiB batches.}}

\label{fig:gz-pipeline}
\end{figure}

\begin{table*}[tp]
  \centering
  \caption{\small{Implementation bases and added source lines across evaluated UMA platforms.}}
  \label{tab:implementation-platforms}
  \scriptsize
  \setlength{\tabcolsep}{3pt}
  \renewcommand{\arraystretch}{1.15}
  \begin{tabular}{@{}>{\raggedright\arraybackslash}p{0.09\textwidth}>{\raggedright\arraybackslash}p{0.28\textwidth}>{\raggedright\arraybackslash}p{0.205\textwidth}>{\raggedright\arraybackslash}p{0.18\textwidth}>{\raggedright\arraybackslash}p{0.18\textwidth}@{}}
    \toprule
    \textbf{Platform} & \textbf{Evaluated stack} &
    \textbf{Userspace interposer (UMD)} &
    \textbf{Page Reclaimer} &
    \textbf{Page Zeroer} \\
    \midrule
    Intel Core Ultra 9 185H &
    Arc Xe-LPG iGPU; Linux 6.16.9; DDR5-5600: 89.6\,GB/s; NVMe: 15.75\,GB/s &
    Built on Intel Compute Runtime 25.12; $\sim$2.7\,K LoC &
    Built in Intel \texttt{i915}; $\sim$1.31\,K LoC &
    Built in Intel \texttt{i915}; $\sim$3.4\,K LoC \\
    AMD Ryzen AI 9 HX 370 &
    Radeon 890M iGPU (gfx1150); Linux 6.16.9; LPDDR5: 120\,GB/s; dual NVMe: 5.45\,GB/s &
    Built on ROCm Systems 7.0.2; $\sim$3.1\,K LoC &
    Built in Linux AMDGPU/KFD; $\sim$1.21\,K LoC &
    Built in Linux AMDGPU/TTM; $\sim$2.6\,K LoC  \\
    NVIDIA Jetson AGX Orin &
    GA10B iGPU (\texttt{sm\_87}); Jetson Linux R39.2 (6.8.12); LPDDR5: 204.8\,GB/s; NVMe: 4.27\,GB/s &
    CUDA 13.2 API interposer; CUDA UMD untouched; $\sim$2.85\,K LoC &
    Built in NVIDIA Jetson Linux driver; 709 LoC &
    Built in NVIDIA Jetson Linux driver; $\sim$2.0\,K LoC  \\
    \bottomrule
  \end{tabular}
\end{table*}

\stitle{Pipelining zeroing tasks.}
To overlap this CPU work with GPU execution, \sys splits each request
into smaller page batches, each processed as one zeroing task.
\sys chooses the per-task size to be the smallest
that does not incur a visible GPU bandwidth drop (1\%), because
extra per-task submission costs can
be overlapped with GPU execution, and smaller batches mitigate bubbles at the start and end of the pipeline.
The three stages then overlap across batches
(Figure~\ref{fig:gz-pipeline}): the CPU prepares and submits task $i+1$
while the GPU executes task $i$, and a separate cleanup worker processes
earlier completed tasks.
Task order is preserved, and completion fences are used to track per-task status.
At \texttt{Finalize}, \sys checks the
completion fences and waits until
the pages are ready.
For the same 4\,GiB request, pipelining reduces latency from 59.76 to
29.64\,ms. However, CPU submission takes 28.8\,ms, exceeding the
25.6\,ms GPU execution time. Preparing the next task can therefore
limit the pipeline's throughput.

\stitle{Reducing CPU overhead.}
First, the Page Zeroer creates a private GPU address
space, reserves GPU virtual addresses, and installs the page-table hierarchy during initialization.
Each task updates existing page-table entries for newly acquired pages
and patches arguments in the command templates. This avoids repeated
page-table and VA allocation and reduces mapping and task cleanup. %
Second, cache maintenance requires sequentially issuing flush operations across the entire memory range.
\sys uses multiple CPU workers to acquire raw pages and process flush operations on independent ranges concurrently,
thus increasing cache-maintenance throughput and shortening the total wall time.

These optimizations reduce CPU submission time below GPU execution
time on both Intel and NVIDIA, reducing gaps between GPU zeroing
tasks. Figure~\ref{fig:gz-pipeline} illustrates this effect on
NVIDIA, where the 4\,GiB request latency decreases from
29.64 to 25.96\,ms.
Detailed measurements and implementation details appear in
\S\ref{sec:eval-aa} and \S\ref{sec:implementation}, respectively.

\section{Implementation}
\label{sec:implementation}

We implement \sys using open-source drivers on Intel and AMD UMA platforms, and extend the implementation to NVIDIA with
a closed-source UMD. Table~\ref{tab:implementation-platforms} lists the modified components and their lines of code (LoC) on each platform.

\stitle{UMD interposer.}
On AMD and Intel, the interposer is integrated into the userspace driver;
the modified APIs are limited to memory management and submission paths
that resolve buffer-object metadata. On NVIDIA it is a
loadable library that interposes the same API classes and uses
\texttt{cuMemAddressReserve} to reserve a private range of GPU virtual addresses for \sys to manage exclusively.
\sys provides two primitive APIs on NVIDIA GPUs for frameworks to mark their model-loading phases,
during which \sys interposes on memory operations. This explicit tagging is necessary
to restrict \sys's independently managed memory range to weight tensors.
Interposing runtime memory allocations can lead to failures when
the CUDA runtime calls APIs that \sys has not implemented through manual implementations.

\stitle{Page Reclaimer.}
The Page Reclaimer similarily implements \texttt{ReclaimPage} across all three platforms. \sys
revokes CPU page-table entries (PTEs) using \texttt{unmap\_mapping\_range}, removes each page-cache
entry and establishes exclusive ownership using \texttt{remove\_mapping}, and
then obtains \texttt{FOLL\_PIN} accounting using \texttt{try\_grab\_folio} manually.
Because \texttt{remove\_mapping} and \texttt{try\_grab\_folio} are not exported to
modules, \sys resolves them through kallsyms lookup.

\stitle{Page Zeroer.}
To coordinate PBOs' delayed physical backing, the Page Zeroer maintains a
demand counter tracking \textsc{Unbound} slots across open PBOs.
\texttt{CreatePBO} increases demand by the requested page count
and returns without waiting for zeroing.
The zeroer prepares private pages in the background, accounting
for ready pages and work in flight.
Successful \texttt{ReclaimPage} calls reduce demand by the
reclaimed page count.
\texttt{Finalize} waits for and installs private pages in the
remaining slots, reducing demand accordingly.
When demand decreases, \sys waits for preparation of any excess pages to finish and releases them.

\sys uses the copy engine on AMD and Intel and the compute engine on NVIDIA;
all three backends are GPU-architecture-specific. AMD's command buffer
 follows the KMD's off-the-shelf code for \texttt{memset} and is submitted directly to
the SDMA ring. Intel's command buffer uses
\texttt{XY\_FAST\_COLOR\_BLT} commands following the UMD specification for userspace \texttt{memset}, and is
submitted as \texttt{i915\_request}s on a KMD-owned blit context.
NVIDIA's compute kernel specializes the Mesa driver's generic compute-fill design
for 64~MiB page batches, using coalesced 128-bit stores and a 64-block grid-stride
launch sized to Orin's 16 SMs and precompiled for \texttt{sm\_87}.
It achieves a speed comparable to \texttt{cuMemset} at sizes larger than 32~MiB
 and is submitted
through a constructed QMD/GPFIFO on a persistent KMD-owned nvgpu TSG/channel in the kernel.
\sys uses 64\,MiB page batches for requests larger
than 64\,MiB and 16\,MiB slots otherwise.
Following the benchmark in \S\ref{sec:eval-aa}, we set a threshold of 8\,MiB for the minimum request size of a zeroing task;
requests smaller than this threshold are handled by CPU zeroing.

\section{Evaluation}
\label{sec:evaluation}
We evaluate \sys's model-loading latency and memory use on three representative UMA platforms, with microbenchmarks
and ablation studies to explain the performance benefits.
Finally, we apply \sys to real application traces to evaluate its performance in practical scenarios.

\begin{figure*}[tp]
    \centering
    \includegraphics[width=\textwidth]{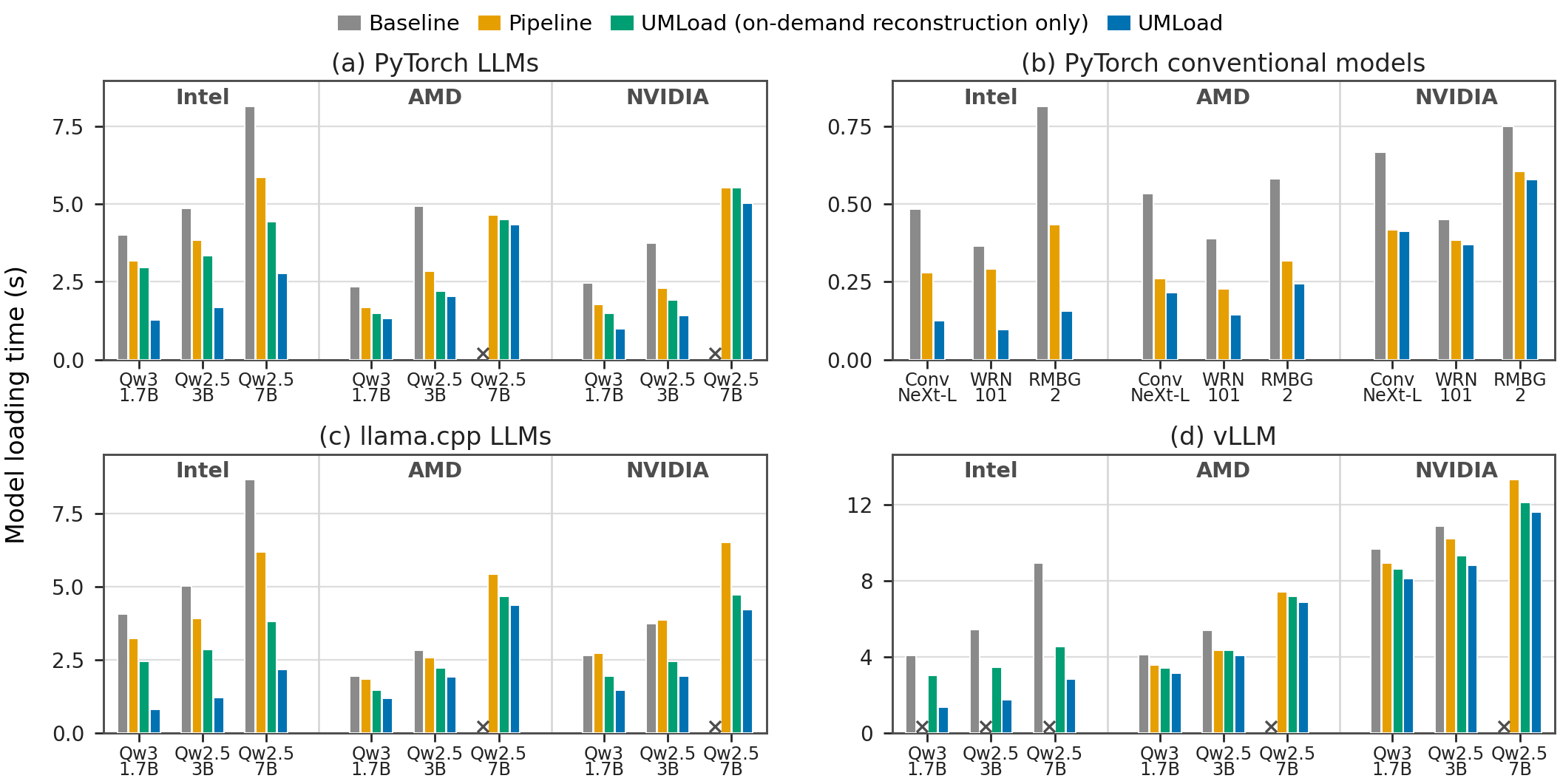}
    \caption{\small{Overall model-loading time on Intel, AMD, and NVIDIA platforms
    for (a) PyTorch LLMs, (b) PyTorch conventional models, (c) llama.cpp
    LLMs, and (d) vLLM.
    Crosses denote unavailable measurements. Lower is
    better.}}
    \label{fig:overall-framework-load}
\end{figure*}

\stitle{Workloads.}
We evaluate across three mainstream inference frameworks: PyTorch (v2.8.0), llama.cpp (tag b8192 for AMD and NVIDIA, b5074 for Intel),
and vLLM (v0.11.2).
Our workloads comprise three LLMs: Qwen3-1.7B~\cite{hf-qwen3-1-7b} (3.2~GiB),
Qwen2.5-3B~\cite{hf-qwen25-3b-instruct} (5.8~GiB), and Qwen2.5-7B~\cite{hf-qwen25-7b-instruct} (13.2--14.2~GiB,
depending on serialization format), and three conventional models:
ConvNeXt-L~\cite{hf-convnext-large} (749~MiB), Wide-ResNet-101~\cite{hf-wide-resnet101} (486~MiB), and RMBG-2~\cite{hf-rmbg2} (845~MiB).
Runtime memory reservation is fixed at 6~GiB for all overall model-loading tests.

\stitle{Baselines and configuration.}
We compare \sys with the following baselines: (1) Vanilla baseline: the vanilla
inference frameworks running on the unmodified kernel and userspace drivers with
the default model loader option.
(2) Pipelined baseline: a framework-level
pipelined loader. As in \S\ref{sec:bg-existing},
for llama.cpp, we use the off-the-shelf
pipelined model loader with direct I/O. For PyTorch and vLLM on AMD and NVIDIA,
we use Run:ai Model Streamer~\cite{runaiModelStreamer2025} to enable
asynchronous file I/O and H2D copies. vLLM on Intel uses the OpenVINO backend and already utilizes a
pipeline that overlaps file I/O and H2D copying.

\stitle{Metrics.}
For LLMs, we measure latency from the first framework loading operation until
prefill begins. For conventional models, we measure from the same
starting point to the completion of weight transfer. Runtime initialization times are excluded.

\subsection{Overall Model-Loading Performance}
\label{sec:eval-overall}
\stitle{Overall latency and memory.}
Figure~\ref{fig:overall-framework-load} compares \sys with vanilla and
pipelined model loading across the three platforms. We report \sys with only
On-demand Reconstruction (ODR) enabled and with both optimizations enabled.
Averaged over the available model--framework configurations, \sys
reduces loading time over vanilla by 71.9\%, 44.7\%, and 36.6\% on
Intel, AMD, and NVIDIA, respectively. The corresponding reductions
over pipelined loading are 62.7\%, 19.9\%, and 22.3\%.
The vanilla Qwen2.5-7B configurations on AMD and NVIDIA are unavailable
because their transient peak memory exceeds the platforms' memory
budgets. \sys supports the model and remains faster than the
available pipelined paths.

The benefit varies across platforms and workloads. Intel generally
shows the largest improvement, and its absolute savings increase
with model size. In PyTorch, the saving over vanilla grows from
2.74~s at 1.7B to 5.38~s at 7B, while the saving over pipelined loading
grows from 1.90~s to 3.09~s. The component measurements explain this
trend: file I/O takes only 64\%--65\% of the combined weight-allocation
and copy time, and the pipeline time beyond file I/O grows from
0.26~s to 1.13~s. Consequently, substantial memory-preparation work
remains exposed even with pipelining, increasing the opportunity
for \sys as model size grows.

On AMD and NVIDIA, the advantage over pipelined loading narrows
for larger models. Averaged over PyTorch and llama.cpp, the reduction
falls from 26.8\% at 3B to 12.8\% at 7B on AMD, and from 44.0\% to
22.1\% on NVIDIA. In the I/O-normalized PyTorch pipeline,
the read component accounts for 58.7\%--59.9\% of pipeline time at
3B, increasing to 97.8\%--98.0\% at 7B. Pipelining therefore leaves
less weight-transfer time exposed for \sys to eliminate. The time saved on
runtime memory allocation is approximately constant across model sizes. Its contribution to relative
speedup consequently decreases as loading time grows.
vLLM additionally retains more model-construction and warm-up work in model loading,
reducing \sys's relative benefit.

For conventional models, \sys outperforms both baselines on Intel
and AMD. NVIDIA's gains over Pipeline are smaller because
its slower disk bandwidth hides more weight transfer than the other two platforms.
For example, ConvNeXt-L takes 412~ms with \sys versus 417~ms with
Pipeline, a 1.2\% reduction; RMBG-2 improves by 4.3\%.

\begin{table}[tbp]
  \centering
  \caption{\small{Peak memory during Intel PyTorch model loading (GiB).
    Values in parentheses show peak memory relative to the weight-file size.}}
  \label{tab:model-load-peak-memory}
  \setlength{\tabcolsep}{4pt}
  \renewcommand{\arraystretch}{1.12}
  \begin{tabular}{lccc}
    \toprule
    \textbf{Method} &
    \textbf{Qwen3-1.7B} &
    \textbf{Qwen2.5-3B} &
    \textbf{Qwen2.5-7B} \\
    \midrule
    Baseline & 6.46 {\scriptsize (2.02$\times$)} & 11.72 {\scriptsize (2.01$\times$)} & 28.60 {\scriptsize (2.01$\times$)} \\
    Run:ai   & 3.91 {\scriptsize (1.21$\times$)} &  6.63 {\scriptsize (1.16$\times$)} & 15.50 {\scriptsize (1.10$\times$)} \\
    \sys     & 3.29 {\scriptsize (1.02$\times$)} &  5.84 {\scriptsize (1.02$\times$)} & 14.27 {\scriptsize (1.01$\times$)} \\
    \bottomrule
  \end{tabular}
\end{table}

We then measure the transient memory
footprint of each loading path.
Table~\ref{tab:model-load-peak-memory} reports peak memory measured with the
cgroup memory controller during Intel PyTorch loading. We subtract growth in
the model-file page cache because it is evictable. The vanilla baseline peaks
at approximately twice the weight-file size, while Run:ai reduces the footprint
to 1.10$\times$--1.21$\times$ by reusing a bounded staging pool. \sys remains
within 1.01$\times$--1.02$\times$ of the weight-file size.
The vanilla baseline's peak memory consists of the weight file and the copy-on-write memory produced
during weight transfer. Run:ai's consists of the weight file and its reusable staging buffer,
while \sys's consists of the weight file and framework-managed memory;
no transient memory amplification is introduced during weight transfer.

\stitle{Ablation.}
We compare \sys with ODR alone and with both optimizations enabled
to quantify their contributions to loading performance.
ODR alone reduces loading time by 10.7\%--55.9\% relative to vanilla
across supported LLM configurations.
It covers 99.9956\%--100\% of weight-transfer bytes; the remaining bytes are in boundary pages handled by copying.
As a result, weight-transfer latency falls to 3.13\%, 7.46\%, and 15.76\%
of its vanilla value on Intel, AMD, and NVIDIA, respectively.

Adding GPU-assisted allocation further reduces loading time
by an average of 1.64~s on Intel, 228~ms on AMD, and 500~ms
on NVIDIA. The differences in absolute savings reflect
each platform's baseline runtime-memory allocation cost.
With both optimizations enabled, weight transfer and
runtime-memory allocation together account for only
9.46\%, 2.93\%, and 13.2\% of total loading time on
Intel, AMD, and NVIDIA, respectively.

\subsection{Effect of On-demand Reconstruction}
\label{sec:eval-cr}

\begin{figure}[tbp]
    \centering
    \includegraphics[width=\linewidth]{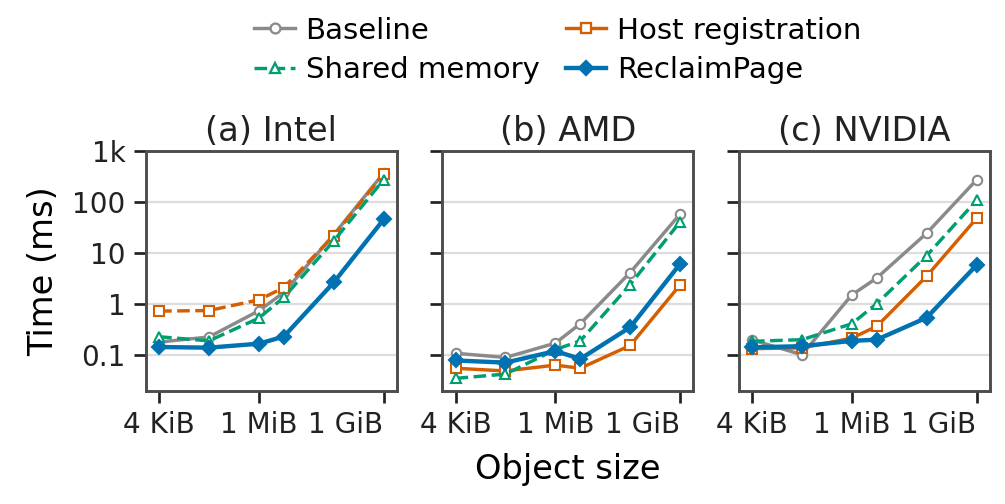}
    \caption{\small{Weight-transfer API latency on (a) Intel, (b) AMD, and
    (c) NVIDIA. Baseline allocates device memory and copies weights
    from the host. Latency excludes file I/O.}}
    \label{fig:micro-pageref-speed}
\end{figure}

\begin{figure}[tbp]
    \centering
    \includegraphics[width=\linewidth]{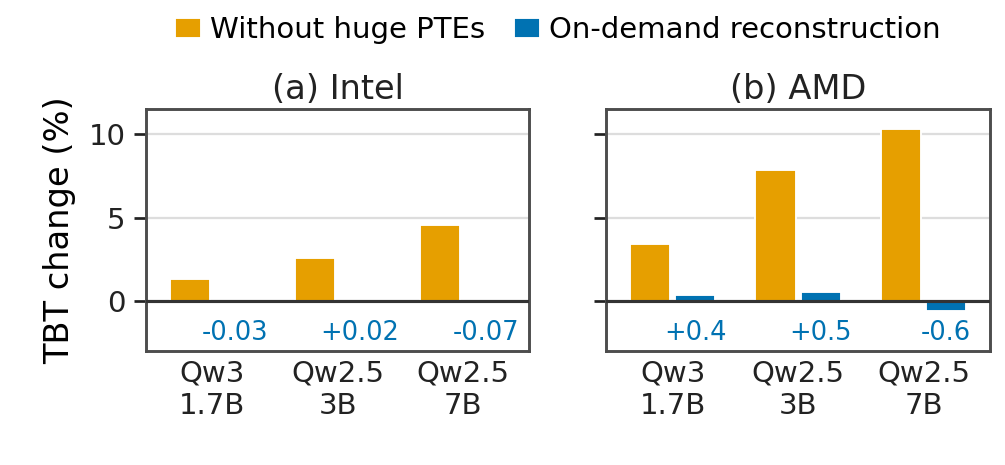}
    \caption{\small{Effect of huge-page mapping recovery on steady-state decode TBT with
    On-demand Reconstruction (ODR) on Intel and AMD. Values show the change relative to vanilla inference.}}
    \label{fig:micro-pageref-recovery-2mstrategy}
\end{figure}

\begin{figure*}[tp]
    \centering
    \includegraphics[width=\textwidth]
        {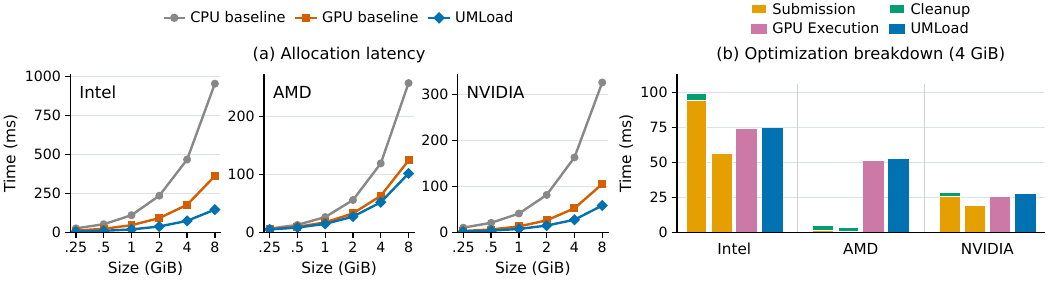}
    \caption{\small{End-to-end allocation latency across three UMA platforms.
    (a) Vanilla allocation using CPU zeroing, baseline allocation using GPU zeroing, and \sys using reduced CPU work and pipelined GPU zeroing.
    (b) \sys's optimization breakdown for 4~GiB. Within each platform, bars from left
    to right show baseline CPU-side latency, \sys CPU-side latency, \sys's GPU
    execution latency, and overall allocation latency using the pipeline
    (\sys).}}
    \label{fig:gpu-zeroing-multiplatform}
\end{figure*}

\stitle{ReclaimPage Latency.}
Figure~\ref{fig:micro-pageref-speed} compares
\texttt{ReclaimPage} API with two zero-copy APIs discussed in \S\ref{sec:bg-existing}
to show \sys's performance relative to alternative weight-transfer APIs.
Note that using the two zero-copy APIs requires framework code to be modified.
The baseline (device allocation followed by an H2D copy) is added as the conventional reference.
The final metric reports the sum of critical steps for each path. Weight-file
I/O is excluded from timing for all methods.

At small sizes, all paths have similar sub-millisecond
latency. At 1~GiB, \texttt{ReclaimPage} is at least 6$\times$ faster than
the baseline and shared memory on all three platforms. AMD host registration is the
only exception.
Host registration only maps the existing source pages, whereas
\texttt{ReclaimPage} takes ownership of the pages and initializes the
corresponding PBO's metadata. On AMD, \texttt{ReclaimPage} spends
4.04~ms committing ownership and another 1.91~ms attaching the pages, while
host registration only changes the GPU page table, attaching the pages in 2.36~ms. This explains why host
registration is faster in its best case (changing the source buffer's mapping to a writable private
mapping will trigger synchronous copy-on-write and severely degrade its performance). In
return, \texttt{ReclaimPage} owns the backing storage,
preserving memory independence and providing more stable and transparent acceleration.

\stitle{Huge-page mapping recovery.}
Figure~\ref{fig:micro-pageref-recovery-2mstrategy} evaluates whether huge pages and
alignment recover decode performance with ODR. Without
rewriting the weight file or enabling THP, reclaimed pages fall back
to 4~KiB entries. This increases time between tokens (TBT) by 1.3\%--4.5\% on Intel and
3.4\%--10.3\% on AMD across the three models.
\sys maintains the proportion of weight bytes covered by huge-page mappings at 96.41\%--99.94\% (vanilla: 100\%).
The residuals are BOs packed with tensors at noncontiguous GPU VAs, for which \sys cannot use \texttt{Reorder}
to restore the contiguity disrupted by earlier \texttt{Padding}.
This recovery also removes the inference regression: TBT with aligned ODR stays
within 0.1\% of the copy-backed baseline on Intel and within 0.6\% on AMD.
Relative to unaligned ODR, alignment reduces TBT by 1.3\%--4.4\% and
3.0\%--9.9\% on Intel and AMD, respectively. The NVIDIA driver uses base-page GPU mappings
and does not exhibit the alignment-induced decode slowdown even without huge pages or 2~MiB realignment (less than 0.5\% change in TBT).

\stitle{Introduced overheads.}
Aligning weights to 2~MiB increases file size
by 1.07\%--5.13\% across the three evaluated Qwen models (ratios are larger on small models).
The file rewrite is performed offline. The larger file
slightly increases the number of bytes read during subsequent model loads, but the padding bytes
are not transferred to the GPU as weights.

Later CPU access can also incur reconstruction I/O, faulting source
pages back in from the file while the GPU retains the reclaimed pages.
For dirty pages, \texttt{ReclaimPage} must complete writeback
before acquiring ownership. In the configured microbenchmark,
median reclamation latency for 1~GiB rises from 41.4~ms to
3.79~s on Intel, from 6.16~ms to 390~ms on AMD, and from
12.7~ms to 330~ms on NVIDIA.
The severe slowdown comes from synchronous
disk operations during writeback.
After reclamation, a CPU access to the source range can fault pages in
from the backing file while the GPU retains the
reclaimed pages. Faulting a cold 5.75~GiB file mapping
takes a median of 717, 1091, and 2091~ms on Intel, AMD, and NVIDIA, respectively,
matching each platform's measured storage bandwidth.
Nevertheless, dirty-page writeback and CPU-side reconstruction are uncommon for
file-backed source buffers in normal loading and were not observed in the workloads we evaluated.

\subsection{Effect of GPU-Assisted Allocation}
\label{sec:eval-aa}

\stitle{Overall allocation latency.}
Figure~\ref{fig:gpu-zeroing-multiplatform}(a) shows that \sys outperforms
both baseline allocation with CPU zeroing and allocation with naive GPU-assisted zeroing
across all three platforms over 256~MiB--8~GiB.
The GPU baseline uses the same in-kernel context and zeroing kernel and
runs the three stages sequentially once.
At 4~GiB, \sys is 2.29$\times$--6.40$\times$ faster
than CPU zeroing and 1.21$\times$--2.40$\times$ faster than the GPU baseline.
Gains between the CPU baseline and GPU baseline come from GPU-zeroing acceleration.
\sys's additional advantage over the GPU baseline comes from pipelining
CPU work with GPU execution and shortening CPU-side work.

\stitle{CPU work reduction and pipelining.}
Pipelining CPU submission, GPU zeroing, and cleanup across chunks reduces
4~GiB allocation latency from 179.41 to 93.92~ms on Intel and from
59.76 to 29.64~ms on NVIDIA.
However, pipelining alone leaves CPU submission as the bottleneck on these platforms.
At 4~GiB, \sys further reduces submission latency from 92.80 to
56.77~ms on Intel and from 28.65 to 17.45~ms on NVIDIA, bringing both below
their GPU completion times of 74.03 and 25.61~ms, respectively.
On Intel, multithreading reduces cache-maintenance wall time
from 78.95\,ms to at most 56.77\,ms.
On NVIDIA, reusing mapping resources
reduces GPU mapping from 10.73 to 1.17~ms and task construction from 1.48 to
0.036~ms.
Together, these optimizations improve pipeline efficiency, reducing allocation
latency further by 20.5\% on Intel and 12.4\% on NVIDIA compared with pipelining alone.
Figure~\ref{fig:gpu-zeroing-multiplatform}(b) shows the resulting CPU-side
latency, including submission and cleanup, against the serial GPU baseline.
AMD's CPU work is already small relative to its GPU phase
(Figure~\ref{fig:motivation-allocation-overheads}). Although CPU-side
time increases from 5.19 to 8.57~ms, \sys reduces allocation latency
from 63.30 to 52.17~ms through its optimized pipeline. Across all
three platforms, the pipeline approaches GPU completion time:
at 4~GiB, end-to-end latency is only 0.9\%, 2.4\%, and 1.1\% above
GPU execution time on Intel, AMD, and NVIDIA, respectively.

\begin{figure}[tbp]
    \centering
    \includegraphics[width=\linewidth]
        {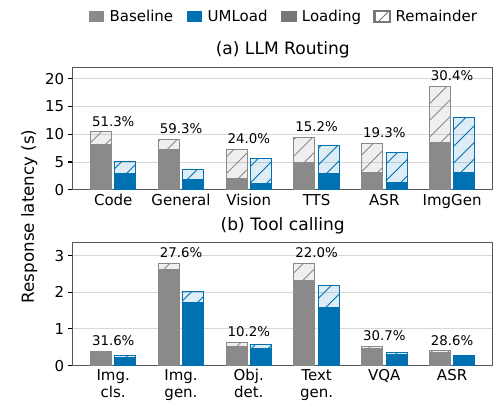}
    \caption{\small{Response latency for (a) LLM routing and (b) tool calling. Code/General measure TTFT;
    multimodal LLMs and tools measure completion time. Percentages give latency reductions.}}
    \label{fig:case-study-results}
\end{figure}

\subsection{Case Studies}
\label{sec:case-study}

We evaluate how \sys reduces response latency in the LLM-routing and
tool-calling workloads introduced in \S\ref{sec:bg-uma}. In both cases,
runtimes remain warm while models are cold-loaded for each request or
tool invocation. We measure TTFT for coding and general requests, and
completion time for multimodal requests and tools.

\stitle{LLM routing.}
We route among six models that cannot all fit in memory:
Qwen3-30B-Int4 for coding~\cite{hf-qwen3-coder-30b},
GLM-4.7-Flash-Int4 for general requests~\cite{hf-glm47-flash},
Qwen3-VL-2B-Instruct (\texttt{Q8\_0}) for
vision-language tasks~\cite{hf-qwen3-vl-2b-instruct-gguf},
Qwen3-TTS-1.7B (F16) for TTS~\cite{hf-qwen3-tts-gguf},
Qwen3-ASR-0.6B (BF16) for ASR~\cite{hf-qwen3-asr-06b-gguf}, and
Flux2 Klein-4B (\texttt{Q4\_0}) for image
generation~\cite{hf-flux2-klein-4b-gguf}.
Request traces interleave samples from
LMSYS-Chat-1M~\cite{zheng-lmsys-chat-1m-2024} and
ToolMMBench~\cite{wang2024toollmm}.
Flux2 and Qwen3-TTS use \texttt{stable-diffusion.cpp} and
\texttt{qwen3-tts.cpp}, respectively; the other models use
\texttt{llama-server}, requiring a 6~GiB reservation for runtime memory. The baseline uses pipelined loading where
supported by \texttt{llama-server} and vanilla loading otherwise;
\sys enables all available optimizations.

Figure~\ref{fig:case-study-results}(a) shows that \sys
improves both types of response latency: TTFT for normal LLMs and end-to-end completion time for multimodal LLMs.
\sys reduces TTFT by 55.3\% on average across coding and general
requests, and completion latency by 22.2\% across the multimodal workloads.
For the latter, model-loading time falls by 49.8\% on average, while
the remainder (mostly inference) accounts for 73.9\% of the optimized response time, showing that
\sys removes much of the startup cost, making the remaining inference and framework overheads
 the dominant remaining cost.

\stitle{Tool calling.}
We replay tool invocations from MNMs benchmark~\cite{ma2024mms} using their
plans, arguments and inputs. The benchmark's PyTorch scripts serve
six local tools: image classification with ViT Base~\cite{hf-vit-base},
image generation with SD-Turbo~\cite{hf-sd-turbo},
object detection with DETR ResNet-101~\cite{hf-detr-resnet101},
text generation with Llama~3.2~1B~\cite{hf-llama32-1b-instruct},
visual question answering with ViLT~\cite{hf-vilt-vqa}, and
ASR with Wav2Vec2 Base 960h~\cite{hf-wav2vec2-base-960h}.

\sys reduces tool response latency by 25.1\% on average, ranging from
10.2\% for DETR to 31.6\% for ViT
(Figure~\ref{fig:case-study-results}(b)).
These gains come from faster loading; inference remains unchanged.
The slight increase in the remaining latency is due to longer disk reads caused by
the larger weight files produced by \sys's rewrite.
ViT has 200 tensors with a median size of 9~MiB, whereas 681 of
DETR's 785 tensors are smaller than 1~MiB.
DETR's many small tensors require more frequent reclamation calls,
and framework-side construction accounts for more of its loading
time, limiting its gains.

\section{Related Work}

\stitle{UMA characterization.}
UMA allows CPUs and GPUs to share physical memory, making duplicate
allocations and data copies avoidable~\cite{smith2024realizing}.
However, exploiting this capability requires appropriate software support.
Wahlgren et al.~\cite{wahlgren-dissecting-2025} characterize allocation,
page faults, address translation, and memory access on MI300A, and
propose porting strategies that address both performance and correctness
when replacing separate CPU/GPU buffers with shared storage.
Yang et al.~\cite{yang2026characterizing} characterize CPU--GPU memory-sharing
correctness and performance across UMA platforms and introduce libumsh,
including a PyTorch case study that eliminates copies by loading weights
directly into shared memory.
Complementary hardware research improves coherence for CPU--GPU
sharing~\cite{nataraja2024coherence}.

\stitle{Application-level UMA adoption.}
Application-level efforts demonstrate UMA's benefits in HPC:
OpenFOAM adopts OpenMP's unified-memory model for GPU
offloading~\cite{tandon2024porting}, iPIC3D uses unified memory for
large-scale plasma simulations~\cite{markidis2025exascale}, and
Quicksilver/CloverLeaf benefit from allocator-aware inter-APU
communication~\cite{schieffer2025interapu}.
These works guide application adaptation to shared memory.
For mobile LLMs, mzCache~\cite{yu2026mzcache} modifies llama.cpp to manage
weights and KV state in shared UMA buffers and overlap restoration of
evicted state with inference. \sys targets model-loader integration,
preserving independent CPU/GPU buffer semantics while eliminating weight
copies, and accelerates zeroing of fresh runtime pages.

\stitle{Runtime support for unified memory.}
Runtime support also enables unified-memory adoption in other settings.
Fumero et al.~\cite{fumeroUnifiedSharedMemory2023} place Java heaps
in unified memory and coordinate garbage collection with GPU execution
to support shared object access.
Bertolli et al.~\cite{bertolli2024zerocopy}
provide implicit zero-copy execution for existing OpenMP applications
and eager GPU page-table mapping to reduce first-access overhead.

\section{Conclusion}

This paper presents \sys, a system that accelerates on-demand model
loading on UMA devices by eliminating redundant weight copies and
accelerating runtime allocation. Our evaluation across platforms,
frameworks, and workloads demonstrates lower loading latency and
memory use, improving response latency for on-device AI.

\balance

{
  \normalsize
  \bibliographystyle{plain}
  \bibliography{bib/references}
}

\end{document}